\documentclass[a4paper,showpacs,twocolumn]{revtex4}
\usepackage{amsfonts}
\usepackage{amsmath}
\usepackage{amssymb}
\usepackage{graphicx}

\begin{document}

\title{Waveguiding properties of surface states in photonic crystals}
\author{A. I. Rahachou and I. V. Zozoulenko}
\affiliation{Department of Science and Technology, Link\"{o}ping
University, 601 74, Norrk\"{o}ping, Sweden}
\date{\today}

\begin{abstract}
We propose and analyze novel surface-state-based waveguides in
bandgap photonic crystals. We discuss surface mode band structure,
field localization and effect of imperfections on the waveguiding
properties of the surface modes. We demonstrate that
surface-state-based waveguides can be used to achieve directional
emission out of the waveguide. We also discuss the application of
the surface-state-waveguides as efficient light couplers for
conventional photonic crystal waveguides.
\end{abstract}

 \maketitle

\textit{Introduction.} Photonic crystals (PCs) have attracted increasing
attention in the last decade due to their unique properties and possible
applications in integrated optical and photonic devices like light emitting
diodes, delay lines, waveguides and lasers \cite{joannopoulos,Sakoda}. Among
variety of photonic-crystal-based devices waveguides play a crucial role not
only as optical interconnections but also as active elements in wide-angle
branches \cite{fan-JOSAB-2001}, channel add/drop filters \cite%
{asano-APL-2003, fan-PRB-1999}, tapered couplers \cite{johnson-PRE-2002},
optical switches \cite{yanik-OL-2003}, etc. Waveguides represent line
defects in periodic crystal structures supporting guided Bloch modes whose
frequency is located in the bandgap. These modes are strongly confined
within the waveguide region and can propagate without loss to substantial
distances. In the present letter we propose a novel type of waveguiding
structures, namely waveguides that operate on surface states of
semi-infinite photonic crystals and \emph{are located on the surface of a PC}%
. Employing surfaces of photonic crystals as waveguides may open up new
possibilities for design and operation of photonic structures for feeding
and redistributing light in PCs.

Surface states reside at the interface between a photonic crystal and open
space, decaying into both mediae \cite{joannopoulos} and propagating along
the boundary. In a square lattice photonic crystal the surface states appear
in the bandgap when a boundary of a PC is modified in some way, by, e.g.,
truncating the surface rods, shrinking or increasing their size, or creating
more complex surface geometry \cite%
{joannopoulos,Mendieta,Zhang,Elson,we-PRB-2005}. The surface modes in a
semi-infinite photonic crystal represent truly Bloch states with the
infinite lifetime and $Q$ factor, and consequently do not couple to
incoming/outgoing radiation. At the same time, it has been demonstrated that
when the translational symmetry along the boundary of the semi-infinite
crystal is broken, the surface mode turns into a resonant state with a
finite lifetime, which can be utilized for lasing and sensing applications
\cite{we-PRB-2005, yang-apl-2004}.
It has also been recently shown that with the help of surface
modes it is possible to achieve directional beaming from the
waveguide opening on the modified surface of a photonic crystal
\cite{moreno-PRB-2004, kramper-PRL-2004}. Surface states there,
coupled with outgoing waveguide radiation, suppress diffraction
and focus outgoing beam. At the same time, so far there have been
no reports on study of guiding properties of PC surfaces.

In order to study surface states in photonic crystals, we apply a
novel computational method based on the recursive Green's function
technique \cite{we-PRB-2005}. The advantage of this method is that
it allows us to calculate and use surface Bloch modes as
scattering states of the system, which makes it possible to
compute the transmission coefficients for surface modes and
corresponding field distributions.

\begin{figure}[htb!]
\begin{center}
\includegraphics[keepaspectratio,width=\columnwidth]{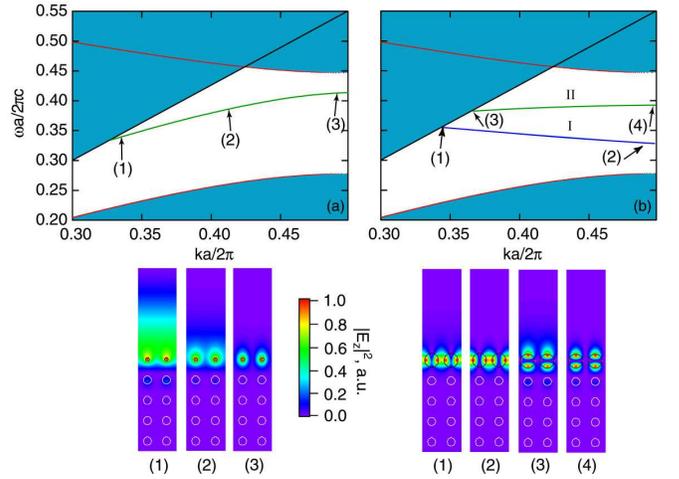}
\end{center}
\caption{(color online) Band structures for TM modes in $\Gamma X$
direction of a square-lattice photonic crystals composed of rods
diameter $D=0.4a$ and permittivity $\protect\varepsilon = 8.9$
along with the projected surface modes. The diameter of the
surface rods is (a) $d=0.2a$ and (b) $d=0.68a$. The fat line
denotes the light line. Lower panel shows the normalized intensity of $E_z$
component in different points of surface-mode dispersion curves.}
\label{fig1-disp-rel}
\end{figure}

\emph{Surface band structure.} We consider a semi-infinite
square-lattice photonic crystal composed of cylinders with
$\varepsilon =8.9$ and diameter $D=0.4a$ ($a$ is a lattice
constant) in air background. We study two different surface
geometries supporting the surface states where the outmost rods
has (a) reduced diameter $d=0.2a$ and (b) enlarged $d=0.68a$. This
photonic crystal has a fundamental bandgap for TM-polarization in
the range $0.33<\omega a/2\pi c<0.44$ and supports one surface
mode for case (a), and two modes for case (b) as shown in Fig.
\ref{fig1-disp-rel}.  The surface modes for these two structures
shows different patterns of field localization. For structure (a)
the field intensity has one maximum within each rod and extends
into the air, quickly decaying into the crystal. For low energies
a significant part of the field intensity extends into a wide
$\sim 5-10a$ air layer near the surface of the PC. This can be
apparently attributed to the proximity between the dispersion
curve of the surface state and the light line, where the group
velocity of the surface state, $v=\partial E/\partial k$, is close
to $c,$ see Fig. \ref{fig1-disp-rel}(a) and Fig.
\ref{fig3-T-inhomogeneties} (a). As the energy increases, the
dispersion curve moves away from the light line, and the field
becomes mainly concentrated on the surface rods. For the case of
the structure (b) with enlarged surface rods, the field is mostly
located within each cylinder and has a node oriented either
horizontally (mode I) or vertically (mode II),see Fig.
\ref{fig1-disp-rel}(b). In contrast with case (a) the intensity of
both surface modes is mainly localized within the surface rods and
its extent to the air is small for the whole energy range.

\begin{figure}[htb!]
\begin{center}
\includegraphics[keepaspectratio,width=\columnwidth]{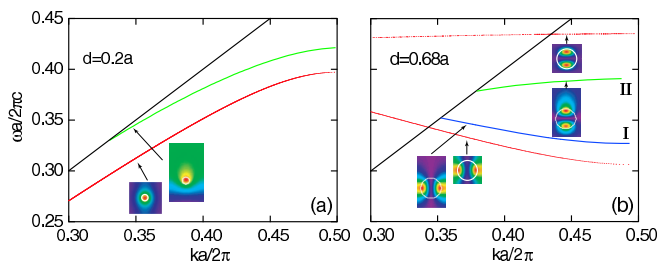}
\end{center}
\caption{(color online) Fragments of the band structures for TM modes in $%
\Gamma X$ direction of infinite square-lattice test photonic
crystals composed of rods with $\protect\varepsilon = 8.9$ and
diameter (a) $d=0.2a$, (b) $d=0.68a$ along with the projected
surface modes of the semi-infinite photonic crystal of Fig.
\ref{fig1-disp-rel}. Field distributions ($E_z$-components) for
corresponding bands are given as the insets.}
\label{fig2-disp-small}
\end{figure}

Let us concentrate now on the surface mode dispersion and the
intensity distribution in structures with reduced and enlarged
surface cylinders. To this end, we construct two test photonic
crystals entirely consisting of corresponding surface rods (i.e.
with diameters $d=0.2a$ and $d=0.68$). Their band structures along
with the projected surface states for the structures (a) and (b)
(shown in Fig. \ref{fig1-disp-rel}) are represented in Fig.
\ref{fig2-disp-small}. The dispersion curve of the surface state
for the
structure (a) begins at the light line and remains nearly linear up to $%
\omega a/2\pi c\gtrsim 0.40$, where its slope slowly decreases and finally
reaches zero. Figure \ref{fig2-disp-small}(a), demonstrate that the shape of
the surface state closely follows the valence band of the test crystal in the $%
\Gamma X$ direction. The same situation holds also for structure
(b), where both surface bands mimic the bulk levels in the
conduction band of the corresponding test PC (given in Fig.
\ref{fig2-disp-small}(b)). Field
distributions for corresponding bands are given as the insets to Fig. \ref%
{fig2-disp-small} and outline the relations between surface-state
bands and corresponding bands of test PCs. It is worth mentioning
that both surface bands for structure (b) have lower velocity in
comparison to structure (a). Fast surface states are known as the
most suitable for waveguidung applications, whereas slow modes can
rather attract interest in structures for "slowing light"
\cite{Figotin_slow_light} or in surface-state cavities
\cite{we-PRB-2005,yang-apl-2004}.

\begin{figure}[htb!]
\begin{center}
\includegraphics[keepaspectratio,width=0.7\columnwidth]{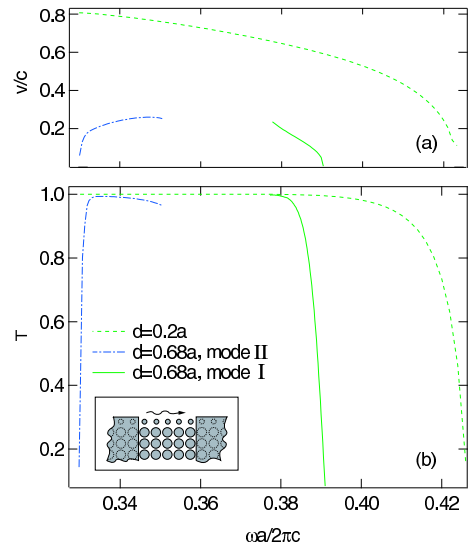}
\end{center}
\caption{(color online) (a) Velocity of different surface modes
from Fig.\protect\ref{fig1-disp-rel}. (b) Transmission coefficient
for surface modes propagating in a non-ideal surface-mode
waveguide. Inset shows the structure under study, where the shaded
regions denote ideal semi-infinite waveguides, and the central
region of the width of 5$a$ represents an imperfect photonic
crystal where scattering of the Bloch surface states takes place.}
\label{fig3-T-inhomogeneties}
\end{figure}

\textit{Effect of inhomogeneities.} Let us focus on the effect of
inhomogeneities of the PC (imperfections in a shape of the rods,
their displacement, or variation of the refraction index
throughout the crystal, etc.) on the waveguiding properties of
surface states. It has been demonstrated previously that such
imperfections strongly affect the performance of lasing
microcavities \cite{we-PRB-microcavities}. We will show below that
such the imperfections can cause a profound impact on the
waveguiding efficiency of the surface modes.

In order to study the effects of imperfections, we divide the system under
study into three regions as shown in the inset to Fig. \ref%
{fig3-T-inhomogeneties}. Two of them are left and right
semi-infinite periodic structures (perfect waveguides for surface
modes), and the block of the PC in between is an imperfect region.
Utilization of the Green's function technique allows us to use
surface Bloch modes as scattering states that propagate in perfect
waveguides from the infinity into the imperfect region where they
undergo scattering. Obviously, in the case when the scattering
region is absent (perfect waveguides are attached to each other),
the Bloch states propagate freely without any scattering.

Because the model is numerical, the discretization of the circular
rods of PCs can not be perfect and thus can be treated as
inhomogeneities or roughness of the structure. Note that the
periodic waveguides are also discretized, but \emph{their
discretization is deliberately chosen differently from that for
the central region}. The central scattering region represents a
photonic crystal of a width of 5 unit cells, each of them being
discretized into 25 meshes ($\sim \lambda /50$) in both $x$ and
$y$ directions. Fig. \ref{fig3-T-inhomogeneties} shows the
velocities of surface states in both structures and corresponding
transmission coefficients. (We note that when the discretization
of each cell in the central region is chosen the same as for unit
cells in the left and right waveguides, the transmission
coefficient through the structure is unity).

The transmission coefficients for each surface mode drop quite
rapidly in the energy regions corresponding to the low velocity of
the surface state. This is because the backscattering probability
is greatly enhanced for the low-speed states. Slow states for
structure with enlarged surface rods are the most strongly
affected. Even for 5 imperfect unit cells the transmission
coefficients for both modes approaches 1 only in very narrow
energy range, which makes these states hardly appropriate for
waveguiding purposes. At the same time the transmission
coefficient for the fast surface state in the structure (a) with
the reduced boundary rods remains 1 in a wide energy region up to
$\omega a/2\pi c\sim 0.40$, which makes it better candidate for
waveguiding applications.

\textit{Applications of surface-state waveguides: Light coupler.}
Due to a unique location on the surface of the PC, surface-state
waveguides can be exploited in a variety of novel applications. In
this letter we focus on two of them, introducing novel light
lead-in structure and sketching the possibility to use a
surface-state waveguide as a directional emitter.

Feeding light into waveguides in photonic crystals composed of
dielectric rods in air background is a complicated challenge, as
normally it requires extremely accurate positioning of a
dielectric waveguide and precise mode matching
\cite{sanches-IEEE-2004, stoffer-OOE-2000}. Even in this case
diffraction at the waveguide termination usually hampers coupling,
and the efficiency of such lead-in systems hardly exceeds 60\%.
Other coupling techniques, like utilization of adiabatic
dielectric tapers \cite{mekis-IEEE-2001} can improve the device
performance but exhibits high sensitivity to parameters of the
tapers.
\begin{figure}
\begin{center}
\includegraphics[keepaspectratio,width=0.6\columnwidth]{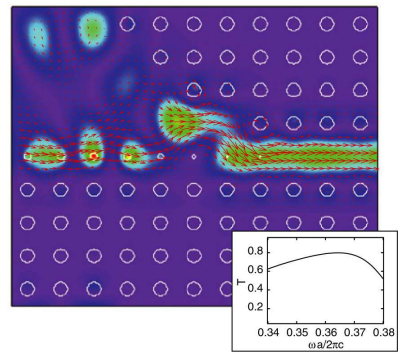}
\end{center}
\caption{(color online) A lead-in coupler structure composed of a
surface-state waveguide to the left and a conventional tapered PC
waveguide to the right. The size of the surface rods gradually
decreases to zero in the central region where the surface-state
waveguide transforms in a conventional PC waveguide. Intensity
distribution is shown for $E_z$-component of the
electromagnetic field at $\protect\omega a/2\protect\pi %
c\approx 0.365$. Arrows depict the flow of the Poynting vector.
Transmission coefficient subject to energy of incoming light is
given in the inset. Parameters of the photonic crystal correspond
to structure (a) in Fig. \ref{fig1-disp-rel}.}
\label{fig:T-leadin}
\end{figure}

In this letter we propose a novel coupler based on waveguiding properties of
surface states.  Figure \ref{fig:T-leadin} illustrates such a lead-in
structure composed of a surface-state waveguide to the left and a
conventional tapered PC waveguide to the right. The diameter of the surface
rods in the surface-state waveguide gradually decreases to zero in the
region of the conventional PC waveguide as shown in Fig. \ref{fig:T-leadin}.
In this device an incoming state in the surface-mode waveguide region enters
a tapered region where it is adiabatically transformed into conventional
waveguiding state.

The maximum achieved transmission reaches $T\approx $ $0.8$, see
inset to Fig. \ref{fig:T-leadin}. We should also mention that
careful optimization of the surface geometry may further improve
the performance of surface-state waveguide couplers, but such the
work is out of the scope of the present Letter.  We also note that
our 2D calculations do not account for the radiative decay in the
direction perpendicular to the plane of the photonic crystal.

\begin{figure}[tbh]
\begin{center}
\includegraphics[keepaspectratio,width=0.6\columnwidth]{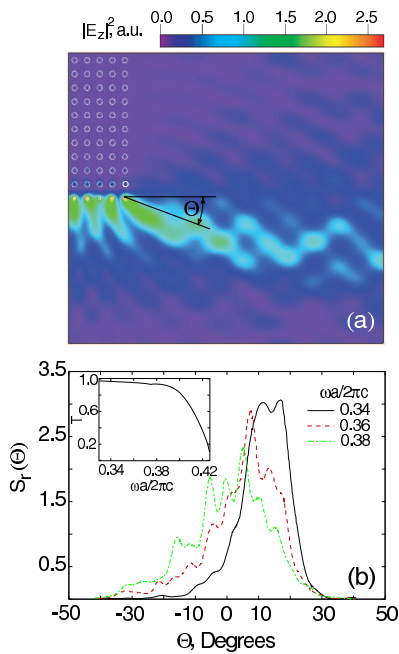}
\end{center}
\caption{(color online)(a) Intensity distribution for
$E_z$-component of the electromagnetic field in the surface-mode
waveguide terminated to air for $\omega a/2\pi c = 0.34$. (b)
Far-field radial component of the Poynting vector $S_r(\theta )$
radiated out of a surface-mode waveguide as a function of
azimuthal angle $\theta$ for different $\omega a/2\pi$. Inset
shows the transmission coefficient for the surface state as a
function of the frequency. Parameters of the waveguide correspond
to structure (a) in Fig. \ref{fig1-disp-rel}. }
\label{fig:T-coupling}
\end{figure}

\textit{Applications of surface-state waveguides: Directional
emitter.} The width of a conventional waveguide in PC is of the
order of wavelength of light $\lambda$.  Because of this, the beam
launched from a semi-infinite photonic crystal into open space is
diffracted at the waveguide opening in a strong angular spread
$\sim 2\pi$. It has been recently shown that it is possible to
achieve directional emission out of a PC waveguides with
corrugated terminations supporting leaky or evanescent surface
states \cite{moreno-PRB-2004, kramper-PRL-2004}. We demonstrate
below that directional emission with the angular spread much less
than in conventional waveguides can also be achieved for
the case of surface-state waveguides coupled to air. Figure \ref%
{fig:T-coupling} shows $E_{z}$ field intensity and directional
diagram for the surface state propagating in semi-infinite
waveguide corresponding to the structure (a) with the surface rods
of a reduced diameter. The most of the beam intensity is localized
within a range $\Delta \Theta \sim 20^{\circ }$. It should be also
noted that the coupling of the surface state to air is rather
high. Inset to Fig. \ref{fig:T-coupling} shows the transmission
coefficient for the surface state in the semi-infinite surface
waveguide propagating into open space. $T$ is close to unity in
the energy region corresponding to high velocity of the surface
state and drops rapidly for energies $\omega a/2\pi c\gtrsim 0.40$
where the velocity of the surface state decreases.

For the case of PC waveguides with corrugated terminations the
directional emission is achieved as the result of the coupling
between the incident beam in the waveguide with the surface modes
on the PC termination that causes the destructive interference for
all directions except a narrow beaming cone
\cite{moreno-PRB-2004,kramper-PRL-2004}. (Note that a mechanism of
the directional beaming in PC waveguides is conceptually similar
to that one in a subwavelength aperture in corrugated metallic
films, where the incident beam is coupled to the surface plasmons
residing at the corrugated boundary \cite{Lezec,Moreno2}).

The origin of a rather narrow beaming cone for the case of
surface-state waveguides is related to the fact that the surface
state is localized in a wide spatial region near the surface
$\lesssim 10a$ (as opposed to conventional waveguides whose width
is typically $\sim a$), see Fig. \ref{fig1-disp-rel}. The angular
spread in this case due to the diffraction, $\sin\theta \sim
\frac{\lambda}{10a}$ is consistent with the calculated far-field
radial distribution of the Poynting vector. As the frequency of
the incoming light increases, the surface mode becomes more
localized, and the spread of the outgoing radiation increases. The
effect of directional beaming in surface-mode waveguides might
find its practical application for integration of PC-based devices
with conventional fiber-optic devices.

\textit{Conclusions} We put forward a novel concept for
waveguiding structures based on surface modes in the bandgap
photonic crystal structures. We analyze surface mode band
structure, field localization and effect of imperfections on the
waveguiding properties of the surface modes. To illustrate
applications of the surface-state waveguides we suggest a new
principle for feeding light into a photonic crystal waveguide and
demonstrate that a semi-infinite surface-state waveguide can be
used as a directional emitter.

\end{document}